\documentclass{JHEP3}
\usepackage{graphicx}
\usepackage{graphics}
\usepackage{cite}
\usepackage{epsfig}
\usepackage{amsmath}
\usepackage{mathrsfs}
\usepackage{amsthm}
\usepackage{amssymb}
\usepackage{oldgerm}
\usepackage{enumerate}
\usepackage{psfrag}
\usepackage{bm,longtable,enumerate}
\usepackage{amsmath,amssymb}
\usepackage[normalem]{ulem}
\def\beq{\begin{equation}}
\def\eeq{\end{equation}}
\def\beqa{\begin{eqnarray}}
\def\eeqa{\end{eqnarray}}
\def\ss{\scriptscriptstyle}

\title{Electromagnetic quasinormal modes of rotating black strings
and the AdS/CFT correspondence}

\author{Jaqueline Morgan$^{\ast}$, Alex S. Miranda$^{\dag}$, 
Vilson T. Zanchin$^{\ddag}$\\
$^{\ast}$Instituto de F\'{\i}sica, Universidade de S\~ao Paulo,\\CP
66318, 05314-970, S\~ao Paulo, SP, Brazil\\
email: {morgan@if.usp.br}\\
$^{\dag}$Departamento de Ci\^encias Exatas e Tecnol\'ogicas,
Universidade Estadual de Santa Cruz,\\
 Rodovia Jorge Amado, Km 16, 45650-000, Ilh\'eus, BA, Brazil\\
email: asmiranda@uesc.br\\
$^{\ddag}$Centro de Ci\^encias Naturais e Humanas,
Universidade Federal do ABC,\\
Rua Santa Ad\'elia 166, 09210-170, Santo Andr\'e, SP, Brazil\\
email: zanchin@ufabc.edu.br}

\abstract{We investigate the quasinormal spectrum of electromagnetic
perturbations of rotating black strings. Among the solutions of Einstein
equations in the presence of a negative cosmological constant there are
asymptotically anti-de Sitter (AdS) black holes whose horizons have the
topology of a cylinder. The stationary version of these AdS black holes
represents rotating black strings. The conformal field theory (CFT) dual of a
black string lives in a Minkowski space with a compact dimension. On the
basis of the AdS/CFT duality, we interpret a CFT plasma moving with respect
to the preferred rest frame introduced by the topology as the holographic
dual to a rotating black string. We explore the consequences of this
correspondence by investigating the electromagnetic perturbations of a black
string for different rotation parameter values. As usual the electromagnetic
quasinormal modes (QNM) correspond to the poles of retarded Green's functions
of $R$-symmetry currents in the boundary field theory. The hydrodynamic
regime of the QNM dispersion relations are analytically studied. Finally, we
investigate numerically the effect of rotation on all the family of
black-string electromagnetic quasinormal modes. We interpret these results
from the CFT perspective and notice the emergence of effects like Doppler
shift of the frequencies and dilation of the thermalization times.}


\preprint{}

\begin{document}

\section{Introduction}
\label{introd}

During the last decade the AdS/CFT correspondence
\cite{Maldacena:1997re,Witten:1998qj, Gubser:1998bc,Aharony:1999ti} at
nonzero temperature and density has become part of the basic toolkit for
those who seek to investigate in- and out-of-equilibrium properties of
systems described by quantum field theories at strong coupling regime. There
is now a variety of bottom-up and top-down AdS/QCD models being used to study
many problems in strong interactions like the computation of deep inelastic
structure functions, the hadronic spectra and the energy loss by heavy quarks
in a plasma (for reviews on the subject see refs.
\cite{BoschiFilho:2006pt,Son:2007vk,Gubser:2007zz,
Erdmenger:2007cm,Iancu:2008sp,Myers:2008fv,Hubeny:2010ry}). More recently,
the AdS/CFT duality has been applied to study condensed-matter systems.
High-temperature superconductivity, quantum and classical Hall effects, and
the (non-)Fermi liquid behavior of some materials are among the physical
phenomena that were investigated by means of holographic methods. An
excellent account of these developments can be found in refs.
\cite{Herzog:2009xv,Hartnoll:2009sz,McGreevy:2009xe}.

Investigations within the AdS/CFT framework have also uncover new aspects of
blackhole physics as the existence of a scalar-hair instability of
Reissner-Nordstr\"om-$\mbox{AdS}_4$ black holes in the regime of low
Hawking's temperature \cite{Gubser:2008px,Hartnoll:2008vx}. Such an effect
opens the possibility of a charged black hole having a superfluid phase, and
the consequent existence of holographic superfluids \cite{Hartnoll:2008kx}.
Another important outcome was the association of hydrodynamics to black hole
physics not only in asymptotically AdS spacetimes, but even for more general
backgrounds. In fact, the first suggestions of applying fluid mechanics to
the study of event-horizon dynamics come from the old `membrane paradigm'
approach
\cite{Thorne:1986iy,Parikh:1997ma,Kovtun:2003wp,Fujita:2007fg,Iqbal:2008by}.
However, this idea has only began to take form with the computation of
transport coefficients via AdS/CFT, and received a great impulse with the
advent of the so-called fluid/gravity correspondence: a one-to-one mapping
between solutions of the relativistic Navier-Stokes equation and the
long-wavelength regime of perturbations of black holes
\cite{Bhattacharyya:2008jc,Bhattacharyya:2008ji,
Bredberg:2010ky,Bredberg:2011jq}. Among some of the interesting results
arising from fluid/gravity duality are the universality of the ratio of the
shear viscosity to the entropy density for all large-$N$ gauge theories with
a dual Einstein gravity \cite{Kovtun:2004de}; the equivalent for static
fluids between area minimization of the fluid and area maximization of the
black hole horizon; and the connection between surface tension of the fluid
and surface gravity of the corresponding black hole \cite{Caldarelli:2008mv}.

The AdS/CFT duality gave rise to a series of results involving traditional
black holes, as well as it allowed new interpretations of blackhole
spacetimes with unusual topology, specially in higher than four spacetime
dimensions. One of the first studied blackhole solutions with a non-spherical
topology in anti-de Sitter spacetime involves a plane-symmetric black hole
solution \cite{Lemos:1994fn,huang,cai}. In addition to the change of the
asymptotic behavior of the blackhole spacetime, the presence or absence of a
cosmological constant determines the different topologies that a black hole
may have. In asymptotically flat four-dimensional spacetimes, a series of
theorems assures that the event horizon of black holes are necessarily
spherical. This is a characteristic of all black holes forming the
Kerr-Newman family. However, the situation changes as soon as we add a
negative cosmological constant term to the Einstein field equations. The
corresponding solutions do not represent globally hyperbolic spacetimes and
the theorems regarding the blackhole topology cannot be applied. It also
allows solutions of the Einstein equations which have all the properties of a
blackhole solution, but with event horizons presenting non-trivial
topologies. Particularly, there are asymptotically AdS spacetimes for which
the spherical horizons of the Schwarzschild-AdS black holes are exchanged by
a plane or a hyperboloid. In the plane-symmetric case, in addition to the
usual planar $\mathbb{R}^{2}$ topology, it is possible to identify points in
the plane surface in order to generate horizons with multiply connected
topologies. The resulting surface can be orientable as the cylinder and the
flat torus or non-orientable like the M\"obius band and Klein bottle
\cite{Brill:1997mf}. In the cylindrical case, the so-called black string (or
cylindrical black hole) can also be put to rotate through a `forbidden'
coordinate transformation in the sense of Stachel \cite{Stachel:1981fg},
which mixes time with angular coordinates and generates new geometries
representing rotating blackhole-like background, i.e., a rotating black
string  \cite{Lemos:1994xp}. The rotating black string is the
object of study in the present work. In the case of the toroidal topology,
the black hole can also be put to rotate, but the analysis is similar and we
do not deal with it here.

The quasinormal modes (QNM) of asymptotically anti-de Sitter black holes have
been intensively studied in the last decade (For recent reviews on QNM, see
e.g. \cite{Berti:2009kk,Konoplya:2011qq}). The main interest in such a
study is the AdS/CFT interpretation in which the electromagnetic and
gravitational quasinormal frequencies of anti-de Sitter black holes are
associated respectively to the poles of retarded correlation functions of
R-symmetry currents and stress-energy tensor in the holographically dual
conformal field theory. According to the dictionary of the duality, a black
hole in the AdS bulk is dual to a CFT thermal state in the AdS spacetime
boundary, and, in particular, as we shall show in the present work, a
rotating black string is the holographic dual to a CFT plasma moving with
respect to the preferred rest frame introduced by the topology.

We investigate the quasinormal spectrum of electromagnetic perturbations of
rotating black strings. Even though the quasinormal oscillations of
rotating black holes with spherical horizon, such as the Kerr-AdS
black holes, have been studied in detail in the literature, the case of
rotating black strings has not being analyzed yet. We then take such a
task and find the dispersion relations for a few values of the
rotation parameter by using the Horowitz-Hubeny numerical method
\cite{Horowitz:1999jd}. Approximate analytical expressions for the
dispersion relations in the hydrodynamic limit are also found.
From the CFT perspective we notice the emergence of
effects like the Doppler shift of frequencies, wavelength contraction and
dilation of the thermalization times.

The layout of the present article is as follows. In the next section, for
completeness, it is reviewed a known plane-symmetric static blackhole
solution of Einstein equations in $(3+1)$ dimensions and a brief summary of
how the rotating black string can be found from the static solution. In the
sequence the metric of the rotating black string is rewritten in a form that
is completely covariant regarding to the spacetime boundary. In section
\ref{EQSperturbations} the electromagnetic perturbation equations are
decoupled in transverse and longitudinal sectors. Analytical results are
obtained and discussed in section \ref{analytical}. Section
\ref{secNumericalQNM} is dedicated to presenting and analyzing the numerical
results and, finally, in section \ref{secfinal} we conclude by commenting
on the main results.

In relation to the units, we are going to use natural units
throughout this paper, i.e., the speed of light $c$, Boltzmann constant
$k_{\ss{B}}$, and Planck constant $\hbar$ are all set to unity,
$c=k_{\ss{B}}=\hbar=1$. Capital Latin indices ($M,\, N,\, ...$) run over
all the spacetime dimensions, while Greek indices ($\mu,\, \nu,\, ...$) run
over the coordinates of the AdS boundary.

\section{The background spacetime}

\subsection{The static black string}
\label{staticblackstring}
 
The background spacetime we start with is a plane-symmetric static
blackhole solution of Einstein equations in $(3+1)$ dimensions. Such a
spacetime presents a physical singularity, and a negative cosmological
constant $\Lambda$, which renders the background geometry asymptotically
anti-de Sitter. In Schwarzschild-like coordinates, the spacetime metric
takes the form \cite{Lemos:1994fn,Lemos:1994xp,huang,cai}
 \begin{eqnarray}
ds^2& =&- \alpha^2r^2f(r)\,d\bar{t}^{\,2} +\dfrac{dr^{2}}{\alpha^2
r^2\,f(r)} +\alpha^2 r^2
\left(dx^{2}+dy^{2}\right), \label{fundo}\\
f(r) &=& 1-\frac{b}{\alpha^3 r^3}, \label{f(r)}
\end{eqnarray}
where $\alpha=1/R=\sqrt{-\Lambda/3}$ and $b$ is a constant related to the ADM
mass of the source. As usual, usual, coordinates $\bar{t}$ and $r$ are
assumed to have the   ranges $-\infty<{\bar{t}}<+\infty$, $0\leq r< +\infty$.
The two-dimensional surfaces ($S$) of constant $\bar{t}$ and constant $r$ may
have different topologies. In addition to the usual topology of a plane
($\mathbb{R}\times\mathbb{R}$), if one or two of the coordinates $x$ and $y$
are periodic the surface $S$ has the topology of a cylinder
($\mathbb{R}\times \mathbb{S}^{1}$) or of a torus
($\mathbb{S}^{1}\times\mathbb{S}^{1}$), respectively (see, e.g.,
\cite{Lemos:2000un} for a review and more references on this subject). Other
multiply connected choices are the M\"obius band and the Klein bottle, but
these are not interesting for the present study because the corresponding
spaces are non-orientable.

In this work we are interested in studying the QNM of rotating AdS black
holes and the interesting cases are the cylindrical black holes
($\mathbb{R}\times \mathbb{S}^{1}$) and the toroidal black holes
($\mathbb{S}^{1}\times \mathbb{S}^{1}$). We consider such cases
because, as it was shown in ref.~\cite{Lemos:1994xp}, these black solutions
allow the addition of angular momentum by coordinate transformations (see,
also \cite{Stachel:1981fg}), generating new spacetimes which represent
rotating black strings and rotating black tori, respectively. On the
other hand, if the dimensions are not compactified the spacetime
\eqref{fundo} can not be put to rotate thereby.

To clarify notation, it is interesting to redefine coordinates as follows:
$\alpha x\rightarrow\bar{\varphi}$ and $y\rightarrow z$, with $0\leq
\bar{\varphi} <2\pi$ and $-\infty<z<+\infty$, for the cylindrical case; or
$\alpha x\rightarrow\bar{\varphi}$ and $\alpha y\rightarrow \bar\theta$, with
$0\leq \bar{\varphi} <2\pi$ and $0\leq \bar\theta<2\pi$, for the toroidal
case. Since the ranges of the angular (compactified) coordinates are
arbitrary, we have chosen these particular ranges for convenience. As
mentioned above, parameter $b$ is related to the mass per unit length $M$ of
the static black string, $b=4M$, with $M$ standing for the linear mass
density, in the cylindrical case, and with $M$ representing the total
gravitational mass of the black hole, in the toroidal case.

The spacetime \eqref{fundo} presents an event horizon at $r ={r_{h}}$,
with
\begin{equation}\label{statichorizon}
 r_{h}=\dfrac{b^{1/3}}{\alpha} = \dfrac{\left(4M\right)^{1/3}}{\alpha}
\end{equation}
 and an essential singularity at $r=0$. 
The Hawking temperature of the black hole is given by
 \begin{equation}\label{hawkingTemp}
{\mathcal{T}}=\frac{3 r_{h}\alpha^2}{4 \pi} = \frac{3 r_{h}}{4 \pi R^{2}},
\end{equation}
where we have put $\hbar =1$.

\subsection{The rotating black string}
\label{rotatingblackstring}

For the sake of clarity in presentation, and to avoid confusion, we restrict
the description to the cylindrical case (black string). The analysis of the
toroidal black hole is easily obtained by further compactifying along the
$y$
direction in metric \eqref{fundo}. The consequences of such an additional
compactification shall not be analyzed in the present work.

According to Stachel \cite{Stachel:1981fg}, a forbidden coordinate
transformation of a static black string (cylindrical black hole) generates
a rotating black string. In this way, Lemos~\cite{Lemos:1994xp} obtained
a rotating black string from the metric given in equation \eqref{fundo}.
We then consider such a solution here by performing the following linear
transformation that mixes the timelike and angular coordinates:
\begin{equation}
\left\{
\begin{aligned}
&\bar{t}=\gamma\left(t-{a}\,\varphi\right),\\
&\bar{\varphi}=\gamma\left(\varphi- a\alpha^2 t\right).
\end{aligned}
\right.
\label{boost}
\end{equation}
After changing from $(x,\,y)$ to $(\bar{\varphi},\,z)$ in \eqref{fundo} and
applying the local transformation \eqref{boost} to the resulting metric one
obtains
\begin{equation}
ds^2= -{\alpha^2r^2}\gamma^2 f(r)\left(dt-ad\varphi\right)^2
+{\alpha^2 r^2} \gamma^2 \left(\dfrac{d\varphi}{\alpha}-{a\,\alpha}
dt\right)^2+{\alpha^2 r^2}dz^2+\frac{dr^2}{\alpha^2r^2 f(r)}\;,
\label{background1}
\end{equation}
where $\gamma$ and $a$ are chosen to satisfy
$\gamma^2\left(1-{\alpha^2}a^2\right)=1$
in order to have the usual anti-de Sitter form of metric at spatial infinity.
Then one has
\begin{equation}
{\gamma}=\dfrac{1} {\sqrt{1-a^2\alpha^2}}. \label{gamma}
\end{equation}
The ranges of the new coordinates are $-\infty<{t}<+\infty$, $0\leq r<
+\infty$, $0\leq {\varphi} <2\pi$ and $-\infty<z<+\infty$, so that
metric \eqref{background1} represents a rotating cylindrical black hole
spacetime.

The stationary metric \eqref{background1}, of course, satisfies Einstein
equations with negative cosmological constant $\Lambda=-3\alpha^2$. Due to the
non-trivial spacetime topology, \eqref{boost} is a
global forbidden coordinate transformation and thus metrics \eqref{fundo} and
\eqref{background1} represent two distinct spacetimes.

As shown initially by Lemos \cite{Lemos:1994xp} and generalized by including
electric charge in ref. \cite{Lemos:1995cm}, for cylindrical topology
($\mathbb{R}\times \mathbb{S}^{1}$), and for $a^2\alpha^2\leq 1$, the metric
\eqref{background1} describes the spacetime of a rotating black string. The
conserved mass $M$ and the conserved angular momentum $J$ per unit length of 
the black string are given in terms of the parameters $a$ and $b$ through
the relations
\begin{equation}
M=b\gamma^{2}\left(\frac{2+a^{2}\alpha^2}{8}\right),\quad\qquad
J=\frac{3ab\gamma^{2}}{8},
\label{mass_angularmomentum} 
\end{equation}
or, inverting for $a$ and $b$, 
\begin{equation}
b=-2\left(M-\sqrt{9M^{2}-8\alpha^{2}J^{2}}\right),\qquad
a=\frac{3M-\sqrt{9M^{2}-8\alpha^{2}J^{2}}}{2\alpha^{2}J}.
\label{paramers_ba} 
\end{equation}
In terms of $M$ and $J$, the condition of existence of an event horizon
$a^2\alpha^2\leq 1 $ becomes $J^2\alpha^2\leq M^2$, and the horizon function
$f(r)=1-b/\alpha^{3}r^3$ reads
\begin{equation}
f(r)=1-\left(3\sqrt{1 -\dfrac{8}{9}\dfrac{J^2\alpha^2}{M^2}}\; -1\right)
\left(\dfrac{2M}{\alpha^3 r^3}\right),
\end{equation}
and in terms of $M $ and $a$ it is
\begin{equation}
f(r)=1-\left(
\dfrac{3\sqrt{4+a^2\alpha^2\left(a^2\alpha^2-4\right)}}{2+a^2\alpha^2}
-1\right)\left(\dfrac{2M}{\alpha^3 r^3}\right) .
\end{equation}

For $0\leq J\leq M/\alpha$, metric \eqref{background1} has a horizon at
$r=r_h$, where $r_h$ is given by
\begin{equation}\label{rh}
r_{h}^{3}=\frac{b}{\alpha^3}=\;2
\frac{\sqrt{9M^{2 } -8J^2\alpha^2 } -M } {\alpha^3} ,
\end{equation}
or 
\begin{equation}
r_{h}^{3}=\frac{b}{\alpha^3}= \left(
\dfrac{3\sqrt{4+a^2\alpha^2\left(a^2\alpha^2-4\right)}}{2+a^2\alpha^2}
-1\right)\left(\dfrac{2M}{\alpha^3}\right) .
\end{equation}

As in the case of the static spacetime \eqref{staticblackstring}, the region
$r=0$ is an essential spacelike singularity, and the region
$r\rightarrow\infty$ is the AdS boundary. The case $J^2=M^2/\alpha^2$
($a^2\alpha^2 =1$) is the extremal limit of the rotating black string, in
which the horizon radius vanishes and the singularity $r=0$ becomes
lightlike. This particular case is of no interest for the present study.

The Hawking temperature of the rotating black string is given
by \cite{Lemos:1994xp}
\begin{equation}
T=\frac{3\alpha}{4\pi}\left[2\left({\,\sqrt{9M^{2}-8J^{2}\alpha^2}-M }\right)
\right] ^ {1/3} \left[\frac{3\sqrt{9M^{2}-8J^{2}\alpha^2}-3M}
{\sqrt{9M^{2}-8J^{2}\alpha^2}+3M } \right ] ^ { 1/2 },
\label{hawkingTemp_rot}
\end{equation}
or, in terms of the horizon radius and the rotation parameter $a$,
\begin{equation}
T=\frac{3r_h \alpha^2}{4\pi}\sqrt{1-a^2\alpha^2}=\mathcal{T}
 \sqrt{1-a^2\alpha^2},
\label{hawkingTemp_rot2}
\end{equation}
where $\mathcal{T}$ is the temperature of a static black string
with the same horizon radius as its rotating counterpart.
For $J=0$ ($a=0$), Hawking temperature \eqref{hawkingTemp_rot2} reduces to
the temperature of the static black string, given by relation
\eqref{hawkingTemp}, as expected.
Moreover, for extremal rotating black strings ($J=M/\alpha$) the Hawking
temperature vanishes, similarly to the case of extremal black holes in
asymptotically Minkowski spacetimes.

Introducing the coordinates $x^\mu =(x^0,x^1,x^2)=(t,\, \varphi/\alpha,\, z)$
and the velocity 
\begin{equation}
U^{\mu}=(U^0,U^1,U^2)=\gamma \left(1,\, a\,\alpha,\,
0\right) \label{velocity}
\end{equation}
on the boundary of the spacetime, i.e., at $r\longrightarrow \infty$, the
metric of the rotating black string \eqref{background1} can be cast into the
form \cite{Bhattacharyya:2008mz,Bhattacharyya:2008jc}
\begin{equation}\label{metric}
ds^2=\frac{dr^2}{\alpha^2 r^2 f(r)}+\alpha^2 r^2\left[-f(r)U_\mu U_\nu 
+h_{\mu\nu}\right]dx^\mu dx^\nu,
\end{equation}
where $h_{\mu\nu}$ is the projector onto the orthogonal direction to $U_\mu$,
\begin{equation}
h_{\mu\nu}=\eta_{\mu\nu}+U_\mu U_\nu, \label {projector}
\end{equation}
and $\eta_{\mu\nu}$ is the Minkowski metric at the boundary,
$\eta_{\mu\nu}=(-1,1,1)$.

The components of the inverse metric tensor $ g^{MN}$ are given by
\begin{equation}
g^{MN} = \left\{\displaystyle{ 
\begin{array}{ll}
g^{rr} =\alpha^2 r^2 f(r), \quad g^{r\mu} = g^{\mu r} =0, \\
 g^{\mu\nu}=\dfrac{1}{\alpha^2 r^2}\left[-\dfrac{U^\mu
U^\nu}{f(r)}+h^{\mu\nu}\right], 
\end{array}}
\right.
\end{equation}
where the indices of the quantities defined at the boundary, such as $U_\mu$
and $h_{\mu\nu}$, are raised (lowered) by the Minkowski metric $
\eta^{\mu\nu}=\eta_{\mu\nu}={\rm diag}(-1,1,1)$. With this notation, we can
find the perturbation equations of a given field for an arbitrary velocity
$U_\mu$.
The case of an electromagnetic field as a  perturbation in the above
spacetime is analyzed and the electromagnetic quasinormal modes are studied
in the following. Even though the geometry of the rotating black string is
locally similar to that of the static black string, the QNM of the rotating
black string are expected to be different from that of the static case
because the QNM depend upon the global structure of the spacetimes, which are
certainly different from each other.

\section{Fundamental equations for the
perturbations}\label{EQSperturbations}

Considering the electromagnetic field as a perturbation on the rotating black
string \eqref{metric}, the resulting equations of motion for the gauge field
$A_{\ss{M}}$ are the usual Maxwell equations
\begin{equation}\label{mov}
\partial_{\ss{M}}[\sqrt{-g}g^{\ss{MN}}g^{\ss{LP}}F_{\ss{NP}}]=0, 
\end{equation}
where $g_{\ss{MN}}$ stands for the metric components given by \eqref{metric},
with capital Latin indices running over all the spacetime dimensions ($M,\, N
= 0, \, 1, \, 2,\,3$), and the Maxwell tensor $F_{\ss{MN}}$ being related to
gauge field $A_{\ss{M}}$ by
\begin{equation}
F_{\ss{MN}}=\partial_{\ss{M}}
A_{\ss{N}}-\partial_{\ss{N}}A_{\ss{M}}.
\end{equation}

We consider fluctuations of the gauge field in terms of Fourier transforms as
follows
\begin{equation}\label{Amu}
A_{\ss{M}}(x^{\mu},r)=\int\frac{dx^{3}}{(2\pi)^{3}}e^{ik_{\mu}x^{\mu}}
\widetilde{A}_{\ss{M} }(k^{\mu},r), 
\end{equation}
where Greek indices run over the coordinates of the AdS boundary, and
$k^\mu$ is the wave vector defined at the spacetime boundary ($\mu, \,
\nu = 0,\, 1,\,2$). The components of the wave vector
along compact spatial directions are in fact quantized numbers. In the
present case, such components are multiples of $\alpha$, the inverse of the
anti-de Sitter radius.

The equations of motion \eqref{mov} for the gauge field \eqref{Amu} together
with appropriate boundary conditions give the QNM dispersion relations for
this field. Since the Maxwell equations are invariant under the gauge
transformation $A_{\ss{M}}\rightarrow A_{\ss{M}}+\partial_{\ss{M}}\lambda$,
we may choose $\lambda$ in such a way to vanish one of the components of
$A_{\ss{M}}$. We adopt the so-called radial gauge, in which $A_r=0$. With
such a choice, we find that the gauge field perturbations satisfy the
following equations
\begin{equation}\label{lr}
k_{\ss{\parallel}}\partial_{r}A_{\ss{\parallel}}-f\,\partial_{r}
(k_{\ss{\perp}}^{\nu}A_\nu)=0,
\end{equation}
\begin{equation}\label{lmu}
\begin{split}
&
\alpha^{4}r^2f\,\left[\partial_r\left(r^{2} f\,\partial_r
A^{\mu}_{\ss{\perp}}\right)-U^{\mu}\partial_r\left(r^{2}\partial_r
A_{\ss{\parallel}}\right)\right]
+ k{\ss{\parallel}}\left(k_{\ss{\parallel}}
A_{\ss{\perp}}^\mu-A{\ss{\parallel}}k_{\ss{\perp}}^{\mu}\right)
\\&+ 
U^{\mu}k^{\sigma}_{\ss{\perp}}\left(k_{\sigma}A_{\ss{\parallel}}-A_{\sigma}k_{
\ss{\parallel}}\right)
-f\,k^{\sigma}_{\ss{\perp}}\left(k_{\sigma}A^{\mu}_{\ss{\perp}}-A_\sigma
k^{\mu}_{\ss{\perp}}\right)=0,
\end{split}
\end{equation}
where, to simplify notation, the tilde was dropped,
$\widetilde{A}_{\mu}\rightarrow A_{\mu}$. The index $\perp$
({\small{\textbardbl}}) indicates the
projection of a given object onto the perpendicular (parallel)
direction to the velocity $U^\mu$. In particular, the longitudinal (parallel)
component of the perturbation gauge field is defined by
$A_{\ss{\parallel}}=U^{\nu}A_\nu$, while the transverse vector is $
A_{\ss{\perp}}^\mu=h^{\mu\nu}A_\nu$.

The system of equations \eqref{lr}--\eqref{lmu} can be decoupled into two
independent (parallel and transverse with respect to the velocity $U^\mu$)
channels. We first project equation \eqref{lmu} onto the parallel direction
to the velocity $U^\mu$, obtaining
\begin{equation}\label{lmul} \alpha^{4}
r^{2}f\,\partial_r\left(r^{2}\partial_r
A_{\ss{\parallel}}\right)-k^{2}_{\ss{\perp}}A_{\ss{\parallel}}
+(k^{\sigma}_{\ss{\perp}} A_\sigma)k_{\ss{\parallel}} =0, \end{equation}
where we have used the identities $k_{\ss{\perp}}^{2}=k_{\ss{\perp}}^{\sigma}
k_{{\ss{\perp}}\sigma}^{}= k_{\ss{\perp}}^{\sigma} k_{\sigma}^{}$. On the
other hand, projecting equation \eqref{lmu} onto the transverse spatial
direction to the velocity $U^\mu$ (by contracting it with $h_{\mu\beta}$) we
have
\begin{equation}\label{lmut}
\alpha^{4}r^{2}f\,\partial_r\left(r^{2}f\,\partial_r
A_{{\ss{\perp}}}^\beta\right)+
\left(k_{\ss{\parallel}}^{2}-f\,k^{2}_{\ss{\perp}}\right)A_{{\ss{\perp}}}^{
\beta }+\left(f\,k^{\sigma}_{\ss{\perp}}A_{\sigma}
-k_{\ss{\parallel}}A_{\ss{\parallel}} \right)
k_{{\ss{\perp}}}^\beta =0.
\end{equation}

The system of independent perturbation equations is now given by equations
\eqref{lr}, \eqref{lmul}, and \eqref{lmut}.

From this point on it is convenient to use normalized quantities. As usual, we
adopt a new radial coordinate $u$ defined as
\begin{equation}
u=\dfrac{b^{1/3}}{\alpha r}=\dfrac{r_{h}}{r},
\end{equation}
with $r_h$ given by \eqref{rh}. Thus, we find
\begin{gather}
\mathfrak{p}_{\ss{\parallel}}\partial_{u}A_{\ss{\parallel}}-f\,\partial_{u}
(\mathfrak{p}_{\ss{\perp}}^{\nu}A_\nu)=0,\label{lrw}\\
f\,\partial^{2}_{u}A_{\ss{\parallel}}-\mathfrak{p}^{2}_{\ss{\perp}}A_{\ss{
\parallel}}+
(\mathfrak{p}^{\sigma}_{\ss{\perp}}A_\sigma)\mathfrak{p}_{\ss{\parallel}}=0,
\label{lmulw}\\
f^2
\partial^{2}_{u}A_{{\ss{\perp}}}^{\beta}+f\left(\partial_{u}f\right)\partial_{u}
A_{{\ss{\perp}}}^{\beta}+\left(\mathfrak{p}_{\ss{\parallel}}^{2}-f\,\mathfrak{p}
^{2}_{\ss{\perp}}\right)
A_{{\ss{\perp}}}^{\beta}+\left(f\,\mathfrak{p}^{\sigma}_{\ss{\perp}}
A_{\sigma}-\mathfrak{p}_{\ss{\parallel}}A_{\ss{\parallel}}\right)
\mathfrak{p}_{{\ss{\perp}}}^{\beta}=0,\label{lmutw}
\end{gather}
where $\mathfrak{p}^{\mu}$ are the normalized wavenumber components defined by
\begin{equation}
\mathfrak{p}^{\mu}=\dfrac{k^{\mu}}{\alpha^{2}r_h} =
\dfrac{3k^{\mu}}{4\pi\,\mathcal{T}}\, , \label{normalized-k}
\end{equation} 
and now $f$ is a function of $u$ given by
\begin{equation}
f=1-u^{3}.
\label{f_u-rotating}
\end{equation}

At this point it is useful to introduce gauge-invariant quantities to be used
as master variables for the perturbation equations. Although there is some
arbitrariness in the choice of gauge-invariant perturbation functions, it is
well known that there are combinations of the gauge field components
$A_{\ss{M}}$ (and their derivatives) which are invariant under the
transformation $A_{\ss{M}}\rightarrow A_{\ss{M}}+\partial_{\ss{M}}\lambda$.
Such combinations include components of the electric field and then it is
interesting to use components of the Maxwell tensor $F_{MN}$ as
master variables. Hence the electric field components, defined by
\begin{equation}
 E_\beta=F_{\beta\nu}U^{\nu},\label{electric-field}
\end{equation}
are taken as master variables. From this, it follows
\begin{equation}\label{ig}
A_{\beta}=\frac{1}{\mathfrak{p}_{\ss{\parallel}}
}\left(\mathfrak{p}_{\beta}A_{\ss{\parallel}}-\frac{E_{\beta}}{\alpha^{
2}r_{h}}\right) .
\end{equation}

Replacing \eqref{ig} into the previous equations \eqref{lrw}--\eqref{lmutw},
we obtain
\begin{equation}\label{lrwE}
\left(\mathfrak{p}_{\ss{\parallel}}^{2}
-f\mathfrak{p}^{2}_ {\ss{\perp}}\right)
\partial_{u}\left(\alpha^{2}r_{h}A_{\ss{\parallel}}\right)+f\partial_{u}
 \left(\mathfrak{p}^{\nu}_{\ss{\perp}}E_{\nu}\right)=0,
\end{equation}
\begin{equation}\label{lmulwE}
f\partial^{2}_{u}\left(\alpha^{2}r_{h}A_{\ss{\parallel}}\right)-\mathfrak{p}^{
\beta}_{\ss{\perp}} E_{\beta}=0,
\end{equation} 
\begin{equation}\label{lmutwE}
\begin{split}
 & f^2
\partial_{u}^{2}\left(E^{\rho}-\alpha^{2}r_{h}A_{\ss{\parallel}}
\,\mathfrak{p}_{{\ss{\perp}}}^{\rho}\right)
+f(\partial_{u}f)\,\partial_{u}\left(E^{\rho}
-\alpha^{2}r_{h}A_{\ss{\parallel}}\,\mathfrak{p}_{{\ss{\perp}}}^{
\rho } \right)
 \\
& +\left(\mathfrak{p}_{\ss{\parallel}}^{2}
-f \mathfrak{p}^{2}_{\ss{\perp}}\right)
E^{\rho}+f\mathfrak{p}^{\sigma }_{
\ss{\perp}}E_{\sigma}\,\mathfrak{p}_{{\ss{\perp}}}^{\rho}=0.
\end{split}
\end{equation}

We now decompose the electric field defined in eq.~\eqref{electric-field}
into its longitudinal and transverse components with respect to the wave
vector, respectively, by 
\begin{equation}
 E^{\sigma}_{\ss{L}}=\frac{\mathfrak{p}^{\nu}_{\ss{\perp }
}E_{\nu}}{\mathfrak{p}^{2}_{\ss{\perp}}}\mathfrak{p}^{ \sigma}_{\ss{\perp}},
\label{EL}
\end{equation}
and 
\begin{equation}\label{ET}
 E^{\sigma}_{\ss{T}}=\left(h^{\sigma\beta}
-\frac{\mathfrak{p}^{ \sigma}_{\ss{ \perp}}
\mathfrak{p}^{\beta}_{\ss{\perp}}}
{\mathfrak{p}^{2}_{\ss{\perp}}} \right)E_{\beta}.
\end{equation}
Using these variables it is possible to combine equations \eqref{lrwE},
\eqref{lmulwE} and \eqref{lmutwE} to obtain the following perturbation
equations 
\begin{equation}\label{eqpertlong}
 f^2\partial_{u}^{2}E^{\sigma}_{\ss{L}}
+\frac{f\;\mathfrak{p}_{\ss{\parallel}}^{2}\;(\partial_{u}f)}
{\mathfrak{p}_{\ss{\parallel }}^{2}- f\mathfrak{p}^{2}_{\ss{\perp}}}
\partial_{u}E^{\sigma}_{\ss{L}} 
+\left(\mathfrak{p}_{\ss{\parallel}}^{2}-f\mathfrak{p}^{2}_{\ss{\perp}}\right)
E^{\sigma}_{\ss{L}}=0,
\end{equation}
\begin{equation}\label{eqperttran2}
f^2\partial^{2}_{u}E^{\sigma}_{\ss{T}}+f(\partial_{u}
f)\partial_ { u } E^ {
\sigma}_{\ss{T}}+\left(\mathfrak{p}_{\ss{\parallel }}^{2}
-f\mathfrak{p}^{2}_{\ss{\perp}}\right)
E^{\sigma}_{\ss{T}}=0.
\end{equation}
Equations~\eqref{eqpertlong} and \eqref{eqperttran2} describe independently
the longitudinal and transverse electromagnetic perturbations of the
rotating black string. For the explicit calculations to find the solutions
of such equations, it is convenient to rewrite them explicitly in terms of
the components of the wave vector with respect to the coordinate basis of
metric \eqref{background1}. More precisely, we use the normalized
quantities $\mathfrak{w}$, $\mathfrak{m}$ and $ \mathfrak{q}$, defined by 
\begin{equation}
(-\mathfrak{w},\alpha\mathfrak{m},\mathfrak{q})=(\mathfrak
{ p}_{0},\mathfrak{p}_{1},\mathfrak{p}_{2}),
\end{equation}
where $m=(4\pi\,\mathcal{T}/3)\,\mathfrak{m}$ may assume only
integer values, since $\varphi$ is an angular (compact) coordinate.
This, together with the velocity $U_\mu$ given by eq. \eqref{velocity}, allows
us to write the perturbation equations \eqref{eqpertlong} and
\eqref{eqperttran2} respectively in the form
\begin{equation}\label{eqpertlong1}
\begin{split}
\partial_{u}^{2}E^{\sigma}_{\ss{L}}+\frac{\gamma^{2}\left(\mathfrak{w}
-a\alpha^{2}\mathfrak{m}\right)^{2}\partial_{u}\left(\ln f\right)}
{\gamma^{2}\left (\mathfrak{w}-a\alpha^{2}\mathfrak{m}\right )^{2}
-f\left[\mathfrak{q}^{2}
+\gamma^{2}\alpha^2\left(\mathfrak{m}-a\mathfrak{w}\right)^{2}\right]}
\partial_{u}E^{\sigma}_{\ss{L}} &+\\
+\frac{\gamma^{2}\left(\mathfrak{w}-a\alpha^{2}\mathfrak{m}\right)^{2}
-f\left [\mathfrak{q}^{2}+\gamma^{2}\alpha^2\left(\mathfrak{m}-
a\mathfrak{w}\right)^{2}\right]} {f^{2}}E^{\sigma}_{\ss{L}}=0,
\end{split}
\end{equation} 
\begin{equation}\label{eqperttran2b}
\partial^{2}_{u}E^{\sigma}_{\ss{T}}+\partial_{u}\left(\ln f\right)
\partial_{u}E^{\sigma}_{\ss{T}} +\frac{\gamma^{2}\left (\mathfrak{w}
-a\alpha^{2}\mathfrak{m}\right)^{2}-f\left[\mathfrak{q}^{2}+\gamma^{2}\alpha^2
\left(\mathfrak{m}-a\mathfrak{w}\right)^{2}\right]}
{f^{2}}E^{\sigma}_{\ss{T}} =0.
\end{equation} 

It is seen that in the case of zero rotation, $a=0$ ($\gamma=1$),
equations \eqref{eqpertlong1} and \eqref{eqperttran2b} become respectively
the polar and axial perturbation equations of the static AdS black hole
presented in reference \cite{Miranda:2008vb}. A comparative analysis  
between the perturbation equations of the static black hole and those of the
rotating black string allows us to establish the following relation between
the frequencies and wavenumbers
\begin{equation}\label{boost2}
\left\{
\begin{aligned}
&\bar{\mathfrak{w}}=\gamma(\mathfrak{w}-a\alpha^{2}\mathfrak{m}),\\
&\bar{\mathfrak{m}}=\gamma(\mathfrak{m}-a\mathfrak{w}),\\
&\bar{\mathfrak{q}}=\mathfrak{q},\\
\end{aligned}
\right.
\end{equation}
where the barred and unbarred quantities refer to the static and rotating cases,
respectively. This suggests that for the
rotating AdS black holes studied here, it is possible to apply the
transformation \eqref{boost2} in the perturbation equations of the static
black string to obtain the perturbation equations for the rotating black
string. Such a procedure should simplify substantially the study of
perturbations of the rotating black string. In particular, we can use all the
formulation developed for the static black hole as long as we consider the
relations between frequencies and wavenumbers given by \eqref{boost2}.
However, let us stress that in the present work we have built all the
perturbation equations directly from the field equations in the rotating
background, without using transformations \eqref{boost2}.

In the next sections we analyze the solutions of equations \eqref{eqpertlong1}
and \eqref{eqperttran2b}  with special interest in the QNM. Let us stress that
\eqref{fundo} and \eqref{background1} really represent different geometries and
therefore we expect to find different quasinormal modes. The boundary conditions
used in the search for such modes are the usual ones for AdS spacetimes, namely,
the condition that there are no outgoing waves at the horizon and the Dirichlet
condition at the AdS boundary.

\section{Electromagnetic quasinormal spectra - analytical
results}\label{analytical}

The perturbation equations \eqref{eqpertlong1} and \eqref{eqperttran2b}
cannot be exactly solved for all frequencies and wavenumbers. However, when
the normalized frequencies and wavenumbers are sufficiently small compared
to unity ($|\mathfrak{w}|,\,|\mathfrak{q}|,\,|\alpha\mathfrak{m}|<<1$), it is
possible to find approximate analytical solutions for the perturbation
equations in the form of power series in $\mathfrak{w}$, $\mathfrak{m}$ and
$\mathfrak{q}$. This procedure is well known in the literature, so that we
do not reproduce it here (see e.g. \cite{Kovtun:2005ev,Miranda:2005qx}). In
this limit we can find the hydrodynamic modes whose main characteristic is
$\mathfrak{w}=\mathfrak{w}_{\ss{R}}-i\mathfrak{w}_{\ss{I}}\rightarrow0$
for $\mathfrak{m}, \mathfrak{q}\rightarrow 0$.

From the point of view of the AdS/CFT correspondence, the study of the
hydrodynamic modes is important because they are the fundamental QNM in
each perturbation sector and therefore they dominate certain processes in
the dual CFT, for instance, the thermalization time of the dual system. The
hydrodynamic dispersion relations have been explored for both the
electromagnetic and the gravitational perturbations of the static AdS black
holes (see refs.
\cite{Herzog:2002fn,Policastro:2002se,Policastro:2002tn,Herzog:2003ke,
Kovtun:2005ev,Miranda:2005qx,Baier:2007ix,Natsuume:2007ty,Miranda:2008vb,
Morgan:2009pn}). Hence, the results found in these works can be used as a
test for the present analysis in the case of small rotation parameter $a$.
In particular, for $a=0$ the dispersion relations found here should imply
in the same results found in the static case.

\subsection{Transverse electromagnetic hydrodynamic mode: not present}

For this perturbation sector there is no solution to eq.~\eqref{eqperttran2b}
satisfying simultaneously the condition of having only ingoing waves at the
horizon and the Dirichlet condition at the AdS boundary, and which are
compatible with the hydrodynamic limit $\mathfrak{m}, \mathfrak{q}
\rightarrow 0$. This implies that, as in the static case, there are no
transverse electromagnetic hydrodynamic QNM for the rotating black string.

\subsection{Longitudinal electromagnetic hydrodynamic mode: the diffusion
mode}

For the longitudinal sector, the procedure to obtain the hydrodynamic QNM can
be simplified. According to the previous discussion, transformations
\eqref{boost2} can be applied to the dispersion relation found for the static
black string. In this way we consider the static dispersion relation 
\cite{Natsuume:2007ty,Miranda:2008vb}
\begin{equation}\label{modo_hidro_ele_2}
\bar{\mathfrak{w}}=-i(\alpha^2\bar{\mathfrak{m}}^{2}
+\bar{\mathfrak{q}}^{2})-i\frac{(9 \ln 3+\sqrt{3}\pi)}{18}
(\alpha^2\bar{\mathfrak{m}}^{2}+\bar{\mathfrak{q}}^{2})^{2}.
\end{equation}
The leading term of this relation corresponds to the dispersion relation of
a diffusion mode, and, after taking into account the normalization factors
(cf. eq.~\eqref{normalized-k}), the diffusion coefficient is
$\bar{D}=3/4\pi\mathcal{T}$.
Considering such a property, we call this analytical approximation as the
diffusion mode.

After applying transformations \eqref{boost2} into the static diffusion
mode \eqref{modo_hidro_ele_2}, we find the
following dispersion relation for the rotating diffusion mode
\begin{equation}\label{elepolhir}
\begin{split}
\gamma(\mathfrak{w}-a\alpha^{2} \mathfrak{m})=&-i\left[
1-\frac{2ia\alpha^2\mathfrak{m}}{\gamma}\left(
1-\frac{2ia\alpha^2 \mathfrak{m}}{\gamma}\right)
+...\right] \left(
\frac{\alpha^2\mathfrak{m}^{2}}{\gamma^{2}}+\mathfrak{q}^{2}\right)\\& -
\frac{i}{18}(-18 a^{2}\alpha^2+\sqrt{3}\pi+9 \ln 3)\left(
\frac{\alpha^2\mathfrak{m}^{2}}{\gamma^{2}}+\mathfrak{q}^{2}\right)
^{2}, 
\end{split}
\end{equation}
which, up to the second order approximation, can be rewritten in the form
\begin{equation}\label{eletranshidr}
 \mathfrak{w}=a\alpha^{2}\mathfrak{m}-\frac{i}{\gamma}\left(\frac{\alpha^2
\mathfrak{m}^{2}}{\gamma^{2}}+\mathfrak{q}^{2}\right)+...\quad
\Longrightarrow \quad \omega=a\alpha^{2}m-\frac{i}{\gamma}\frac{3}{4\pi
\mathcal{T}}\left(\frac{\alpha^2 m^{2}}{\gamma^ {2}}+q^{2}
\right)+...
\end{equation}
where the ellipses denote higher powers in $\mathfrak{m}$ and
$\mathfrak{q}$. The last relation is then an analytical approximation for
the dispersion relation of the hydrodynamic (electromagnetic) QNM
of a rotating black string.

Expression \eqref{eletranshidr} helps to clarify the analysis. Initially
we observe that the diffusive mode, differently of static case, is not purely
damped, i.e., the quasinormal frequencies have nonzero real parts. The first
order approximation real term ($a\alpha^{2}\mathfrak{m}$) can be interpreted
as a convection term that emerges because of the motion of dual fluid with
respect to the inertial observer. We can also see that the second term, i.e.,
the leading term of the imaginary part of the frequency, is a diffusion term.
From such a term we can read the diffusion coefficient $D=3/\gamma4\pi
\mathcal{T}$ and immediately observe that it differs from the diffusion
coefficient of the static case by a $\gamma$ factor (cf. the non-normalized
version of eq. \eqref{modo_hidro_ele_2}). The diffusion coefficient is a
global factor in the imaginary part of the frequency that determines the
characteristic damping time of a QNM, given by $\tau=1/\omega_{\ss{I}}$.
According to the AdS/CFT duality, such a characteristic damping time is
related to the thermalization time, i.e., to the timescale that the perturbed
dual thermal system spends to return to thermal equilibrium. For $m=0$, our
results allow us to write
\begin{equation}
\tau=\gamma\frac{4\pi \mathcal{T}}{3q^2},
\end{equation}
and since $\bar{\tau}=4\pi \mathcal{T}/3q^2$, we have 
\begin{equation}\label{tau}
\tau=\gamma\bar{\tau}.
\end{equation}
The above result indicates dilation of damping times of the QNM of the
rotating black string (and of the thermalization times of the dual thermal
system) when compared to the static case. 

The relation between the wavenumber $m$ of the QNM of the rotating black
string and the corresponding wavenumber of the static black string $\bar m$
is very simple, namely $m=\gamma\bar{m}$. Taking into account that the
wavenumber $m$ is in the direction of rotation, this is a reasonable result
and is interpreted as a length contraction due to the motion of the black
string, or due to the motion of the dual fluid in the CFT side.

Although the analytical approximation \eqref{elepolhir} is not a purely
diffusive mode (it has a real term), we adopt the same
nomenclature as for the static case and call this approximation the diffusion
mode.

As expected, the result of equation \eqref{elepolhir}
reduces to the static hydrodynamic mode in the limit of zero rotation,
$a\rightarrow0$ ($\gamma\rightarrow1$)
\cite{Natsuume:2007ty,Miranda:2008vb}. This hydrodynamic mode was also
studied numerically as we shall see in section \ref{secNumericalQNM}.

\section{Electromagnetic quasinormal spectra - numerical results}
\label{secNumericalQNM}

In this section we use the numerical method developed by Horowitz and
Hu\-beny \cite{Horowitz:1999jd} to obtain the quasinor\-mal frequencies of
the $(3+1)$-dimensional rotating AdS black string. This method consists in
a series expansion of the considered perturbation function, and transforms
the problem of solving a differential equation into the problem of finding
roots of a polynomial (see also \cite{Morgan:2009pn} for more details). The
numerical results are presented and discussed and whenever possible they are
compared to the analytical results, and interpreted in terms of the dual
thermal system.

\subsection{Hydrodynamic quasinormal modes}

The hydrodynamic QNM are those for which the frequency
$\mathfrak{w}(\mathfrak{m},\mathfrak{q})$
vanishes when $\mathfrak{m},\mathfrak{q}\rightarrow0$, and, as discussed
above, for electromagnetic perturbations these modes arise only in the
longitudinal sector of perturbations.

\subsubsection{Wavemodes propagating perpendicularly to the rotation
direction}

The solid lines in the graphs of figure \ref{hidro} show two dispersion
relations for the electromagnetic hydrodynamic QNM with $\mathfrak{m}=0$,
i.e.,  with the component of the wavenumber parallel to the rotation
direction fixed to zero, meaning that such modes propagate along the
perpendicular direction with respect to the rotation direction.
 As seen above, the hydrodynamic quasinormal mode corresponding to a given 
value of $a$ is a purely damped mode. For comparison, the analytical
approximations (diffusion mode), obtained from eq.~\eqref{elepolhir} with
$\mathfrak{m} =0$, are also shown. We find a good agreement between numerical
(hydrodynamic QNM) and analytical (diffusion mode) results in the
hydrodynamic regime.

\FIGURE{
\centering\epsfig{file=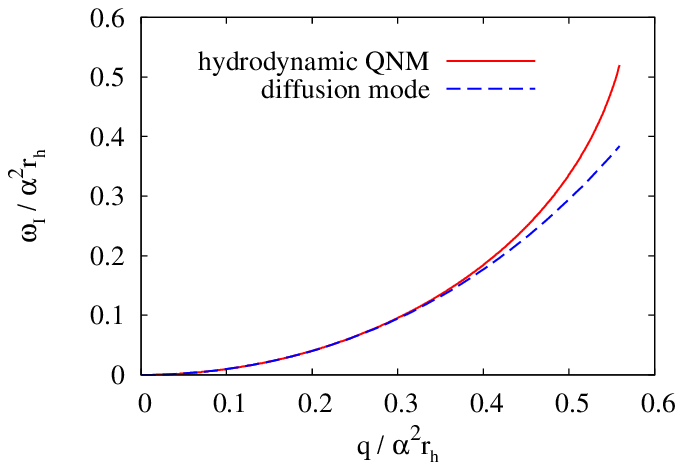,height=7.137cm,width=4.968cm,angle=270}
\centering\epsfig{file=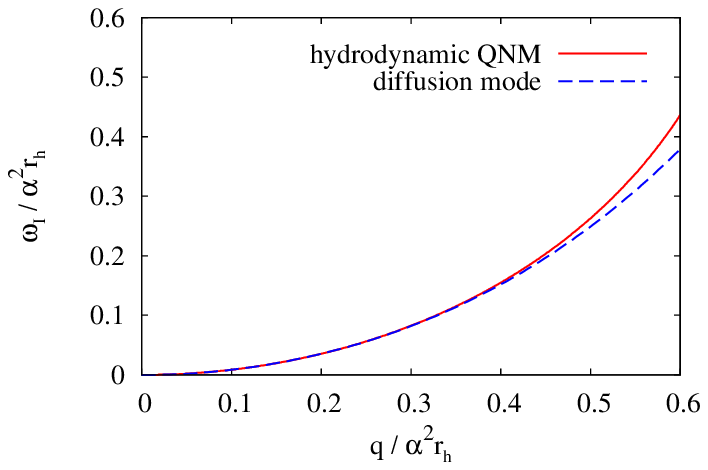,height=7.137cm,width=4.968cm,angle=270
}
\caption{The dispersion relations for the electromagnetic hydrodynamic
quasinormal modes of the rotating black string with $\mathfrak{m}=0$ for
$a\alpha=0.2$ 
(left) and $a\alpha=0.5$ (right). The solid (red) lines show the numerical
results, while the dashed (blue) lines are plots for the diffusion mode
\eqref{elepolhir}. Solid lines terminate at the indicated points, while
the dashed lines extend to higher values of $q/\alpha^{2}r_h$. As
expected, the analytical results (diffusion mode) are in good agreement
with the numerical results (hydrodynamic QNM) just in the hydrodynamic
limit ($\mathfrak{w}\ll1$ and $\mathfrak{q}\ll1$).} \label{hidro}
}

The behavior of the hydrodynamic QNM propagating along the perpendicular
direction to the rotation, i.e., for which $\mathfrak{m}=0$, and for small
rotation parameters $a$,  is similar to the static case. In particular, there
exist a saturation point, i.e., a maximum wavenumber value
$\mathfrak{q}_{\mbox{\scriptsize{lim}}}$, beyond which the specific mode
disappears (for more details see ref. \cite{Miranda:2008vb}, and
for similar QNM emerging in other spacetimes, see ref. \cite{Myers:2008me}).
This implies that both of the solid lines in the graphs of figure~\ref{hidro}
terminate at
the largest values of the wavenumber $\mathfrak{q}= q/\alpha^{2}r_h$ showed
in the figure. However, as the rotation parameter increases the point of
saturation moves to higher wavenumber values, and finally, for some critical
value of the rotation parameter $a_c$, it disappears and the mode extends to
arbitrarily large values of $\mathfrak{q}$. The graphs of figure
\ref{hidropuram} exemplify this behavior. One can see that for $a\alpha=0.1$
the point of saturation occurs in $\mathfrak{q}\simeq0.561$, and for
$a\alpha=0.3$ in $\mathfrak{q}\simeq0.595$. On the other hand, for
$a\alpha=0.5$ and $a\alpha=0.7$ the point of saturation disappears and the
purely damped dispersion relation extends continuously for higher values of
$\mathfrak{q}$. A more detailed numerical analysis indicates the critical
value $a\alpha=a_{c}\,\alpha= 0.395$ as the limit where there still exists a
saturation point (in $\mathfrak{q}\simeq0.633$), but for  $a\alpha \gtrsim
0.395$ the point of saturation no longer exists. 
This peculiar feature is still not well understood since the
saturation point is outside the hydrodynamic limit and the analytical
expressions are not 
valid in this region. However, we find a similar behavior in other 
physical systems, like the D3-D7' model in the presence of
a background magnetic field and at finite density \cite{Jokela:2012vn}.
In figure \ref{hidropuram}, we can see again a good agreement between all the
analytical and numerical results in the hydrodynamic limit.

\FIGURE{
\centering\epsfig{file=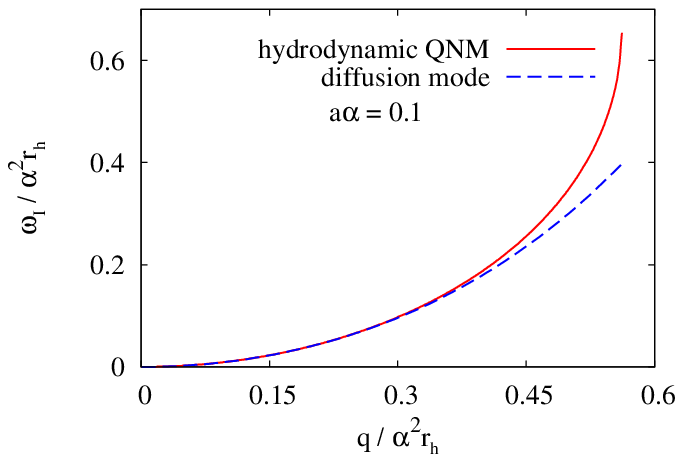, height=7.14cm, width=4.97cm,
angle=270}
\centering\epsfig{file=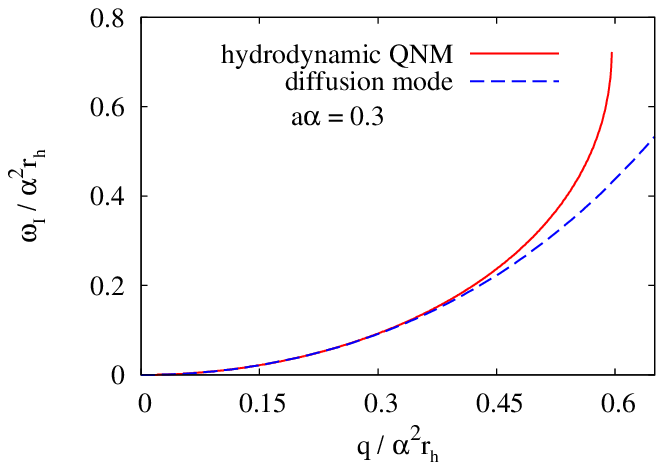, height=7.14cm, width=4.97cm,
angle=270}
\centering\epsfig{file=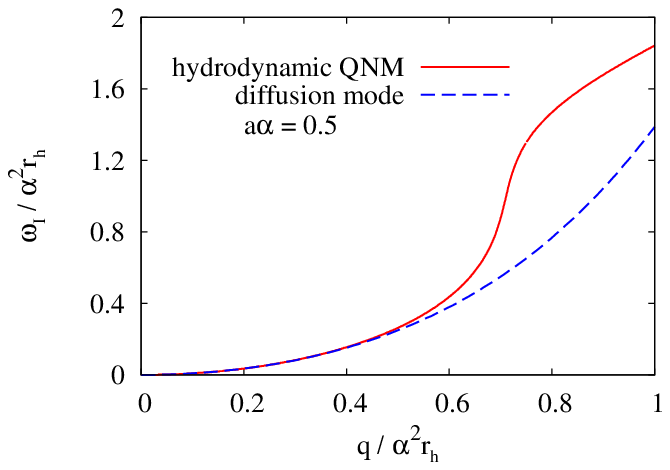, height=7.14cm, width=4.97cm,
angle=270}
\centering\epsfig{file=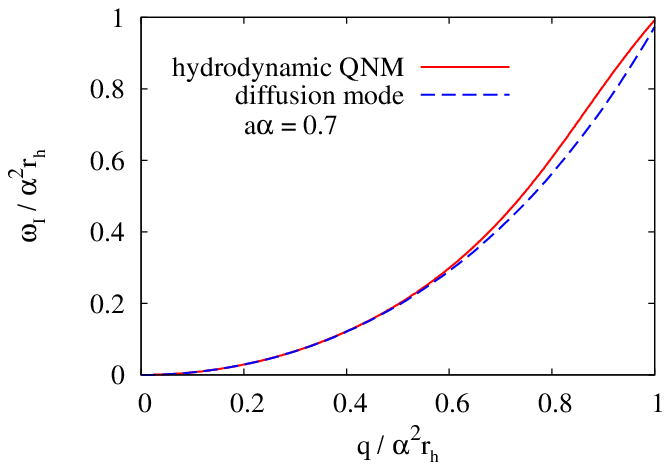, height=7.14cm, width=4.97cm,
angle=270}
\centering\caption{Dispersion relations for the special case of the
hydrodynamic quasinormal mode with $\mathfrak{m}=0$ (purely damped). As in the
static case, for $a\alpha=0.1$ and $a\alpha=0.3$ there is a saturation point beyond
which the hydrodynamic QNM ceases to be purely damped, but for $a\alpha=0.5$
and $a\alpha=0.7$ the purely damped mode extends to higher values of
$\mathfrak{q}$.}
\label{hidropuram}
}

\subsubsection{Wavemodes propagating along the rotation direction}

We also analyzed the hydrodynamic quasinormal modes propagating along the
rotation direction $\varphi$, i.e, modes with wave-vectors parallel to
rotation direction, characterized by wavenumber of the form $\mathfrak{q}=0$
and $\mathfrak{m}\neq 0$. In this case, differently from the static case and
differently from the rotating case for $\mathfrak{m}=0$, the dispersion
relations have nonzero real and imaginary parts (see figure \ref{figele2})
for all wavenumber values. One of the reasons for the emergence of a non-zero
real part for the frequency is the convective motion of the dual fluid with
respect to the preferred rest frame introduced by the topology. Again we can
see a satisfactory agreement between analytical and numerical results in the
hydrodynamic regime.

\FIGURE{
\centering\epsfig{file=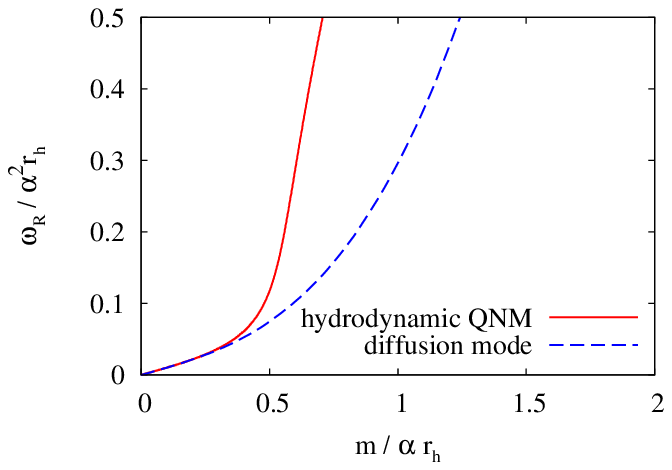, height=7.137cm,
width=4.968cm, angle=270}
\centering\epsfig{file=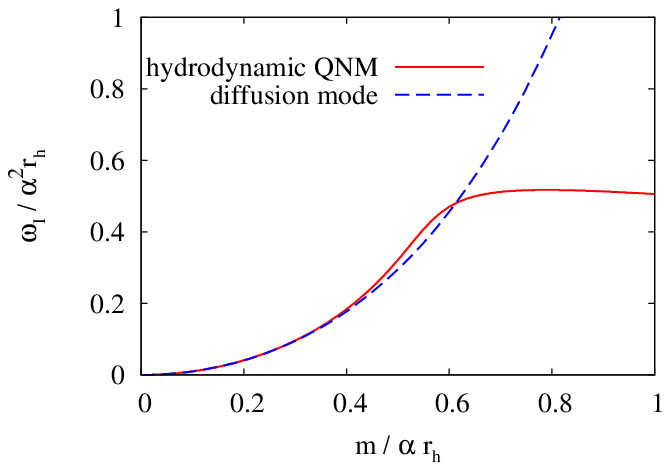, height=7.137cm,
width=4.968cm, angle=270}
\caption{The dispersion relations
for the electromagnetic hydrodynamic QNM of a rotating black string with
$\mathfrak{q}=0$ and $a\alpha=0.1$. Both the real (left) and the imaginary (right)
parts of the quasinormal frequency are shown. The solid lines correspond to
the hydrodynamic quasinormal mode and the dashed lines correspond to the
diffusion mode \eqref{elepolhir}. Again, we can see the agreement between
numerical and analytical results in the hydrodynamic limit
$(\mathfrak{w},\mathfrak{m})\rightarrow0$, as expected.}
\label{figele2}}

\subsection{Purely damped quasinormal modes - slow rotation}

Purely damped non-hydrodynamic QNM are also found in the electromagnetic
perturbations of rotating black strings. These modes arise both in the
longitudinal and in the transverse sectors for perturbations propagating only
along the perpendicular direction to the rotation (with $\mathfrak{m}=0$).
The behavior of the dispersion relations of such modes is similar to static
case \cite{Miranda:2008vb}. The purely damped modes of the static AdS black
hole are found only for small values of wavenumber $\mathfrak{q}$ (for
$\mathfrak{m}=0$), and extend to high values of the frequencies in a tower of
modes with equally spaced imaginary frequencies. However, the higher is the
frequency, the smaller is the domain of values of $\mathfrak{q}$ where the
modes exist. As obtained from the numerical study, the behavior of the purely
damped modes of the rotating black strings strongly depends on the value of
the rotation parameter $a$. These modes are not equally spaced in the
frequency, and disappear above some critical value of the frequency, which
depends on the rotation parameter. On the other hand, for large values of
$a$, it appears a different
class of purely damped modes which extend to high values of the wavenumber.
All of these modes were studied in some detail, as shown in figures
\ref{modospur} and \ref{modospur2}. 

Figure \ref{modospur} shows some of the typical
results found, where we can see that at small values of the $a$, the
purely damped modes have a standard behavior that is repeated for different
values of the rotation parameter. We discuss here the behavior
of dispersion relations for $a\alpha=0.1$ (center panels of figure
\ref{modospur}). For the other values of the $a$ the analysis is similar.

\FIGURE{
\centering\epsfig{file=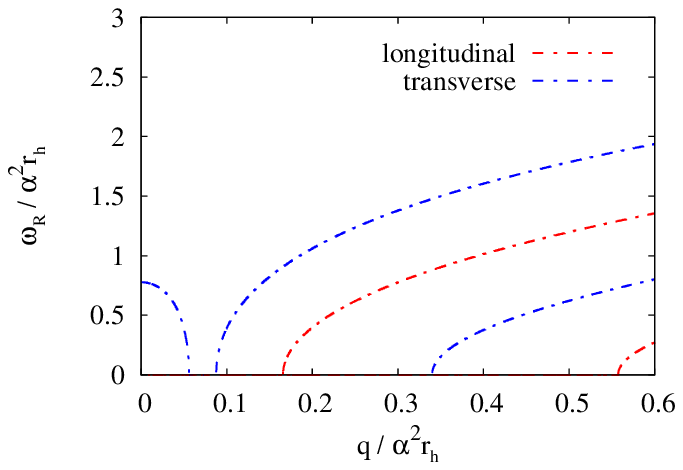, height=7.14cm, width=4.97cm,
angle=270}
\centering\epsfig{file=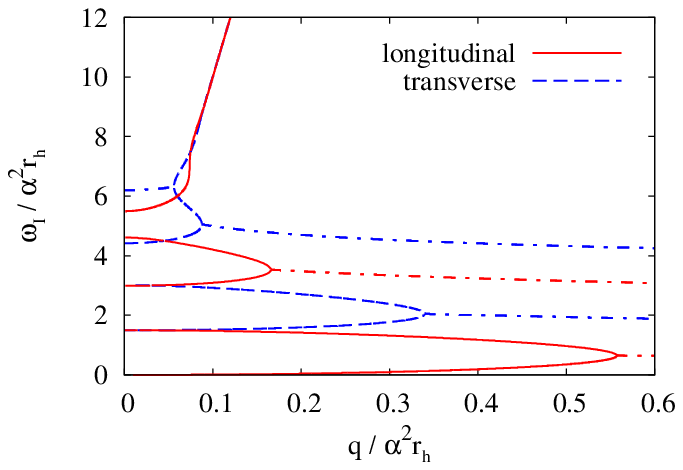, height=7.14cm, width=4.97cm,
angle=270}
\centering\epsfig{file=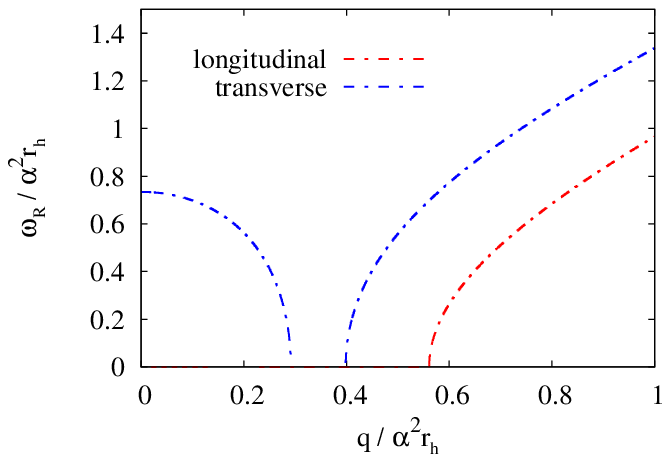, height=7.14cm, width=4.97cm,
angle=270}
\centering\epsfig{file=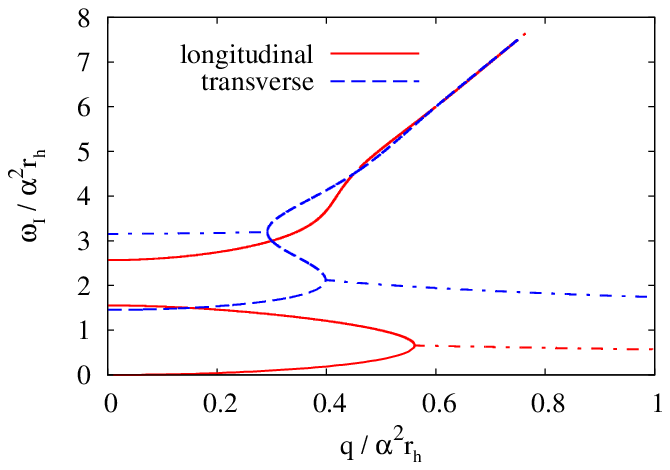, height=7.14cm, width=4.97cm,
angle=270}
\centering\epsfig{file=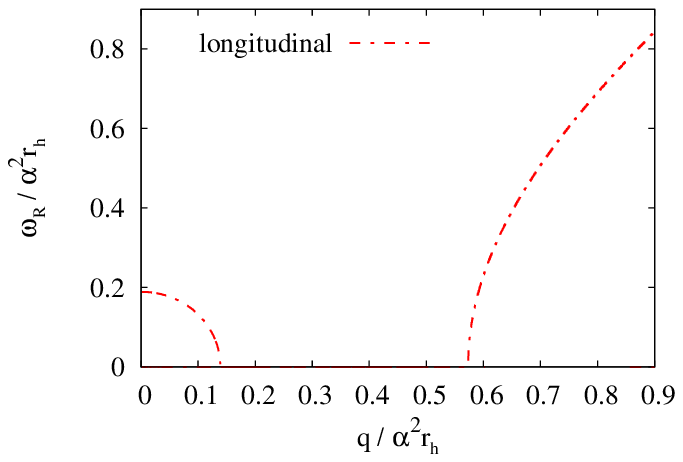, height=7.14cm, width=4.97cm,
angle=270}
\centering\epsfig{file=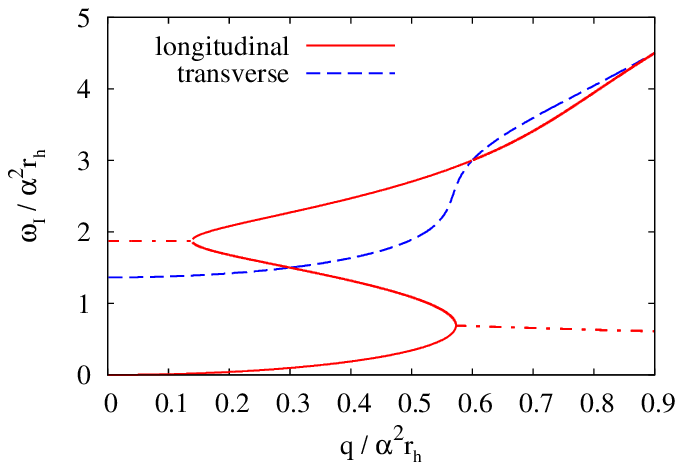, height=7.14cm, width=4.97cm,
angle=270}
\caption{The dispersion relations for the purely damped QNM are shown
by the solid and dashed lines (plots on the right) for different values
of the rotation parameter. From top to bottom, we have $a\alpha=0.01$,
$a\alpha=0.1$ and $a\alpha=0.2$. For completeness, ordinary modes
(dot-dashed lines) connected to the purely damped ones are also shown. Real
(imaginary) parts of the frequencies of these modes are shown in the graphs
on the left (right) hand side. }
\label{modospur}
}

For $a\alpha=0.1$, we found just three longitudinal purely damped modes (red
solid lines), where the hydrodynamic QNM and the first purely damped
non-hydrodynamic mode do not show new features in comparison to the static
case \cite{Miranda:2008vb}. There is a saturation point in
$\mathfrak{q}\equiv
\mathfrak{q}_{\ss{lim}}\sim0.56$, where the frequencies of the two modes
coincide, and for wavenumber values larger than $\mathfrak q_{\ss{lim}}$
the purely damped modes no longer exist. As in the static case, for larger
values of $\mathfrak{q}$ there is an ordinary QNM, i.e., a mode whose real
and imaginary parts of the frequency are both nonzero, and that starts
exactly at the saturation point (see the dot-dashed lines in figure
\ref{modospur}). On the other hand, the third longitudinal mode does not
present saturation point and, for all values of the $\mathfrak{q}$ we have
investigated, it remains purely damped. Differently from the static case
where it exists an infinite tower of purely damped modes for small wavenumber
values, no other purely damped longitudinal mode was found. The imaginary
part of the frequency of the last purely damped longitudinal mode found grows
very fast with the wavenumber, meaning that such modes are highly damped.
This is a particular characteristic of the purely damped QNM of rotating
black strings.

For transverse perturbations, and with $a\alpha=0.1$, we also find just three
purely damped modes (dashed blue lines in figure~\ref{modospur}). The
first mode, similar to the static case,  begins at $\mathfrak{q}=0$, extends
to $\mathfrak{q}\sim0.39$, disappearing for $\mathfrak{q}>0.39$. The second
purely damped mode is present only in the region
$0.29\lesssim\mathfrak{q}\lesssim0.39$, and it meets the first mode at
$\mathfrak{q}\sim 0.39$. Finally, the third purely damped QNM is identified
as belonging to the region $\mathfrak{q}\gtrsim0.29$ and, as the wavenumber
increases, this purely damped mode extends to higher values of the frequency.
For the values of $\mathfrak{q}$ considered in the analysis, the third mode
does not present real part of the quasinormal frequency.  {It is also seen
that the third transverse mode grows fast with the wavenumber, merging with
the third longitudinal mode in the limit of high wavenumbers}. As seen
in the second graph in the right panel of figure \ref{modospur}, for
wavenumber values lower than the saturation point, the behavior of the first 
purely damped mode is similar to the static case. 
However, the second purely damped mode does not exists for 
$\mathfrak{q}\lesssim 0.29$. Moreover, another important difference when
compared to the static case is the presence of an ordinary mode, a mode whose
frequency has nonzero real and imaginary parts in the region of small
wavenumbers ($0\leq\mathfrak{q}\lesssim 0.29$). In fact,
this kind of ordinary mode is present only for small wavenumbers, where the
imaginary part of the frequency is approximately a constant, and it touches
the last purely damped mode found for each sufficiently small given value of
the rotation parameter.

\FIGURE{
\centering\epsfig{file=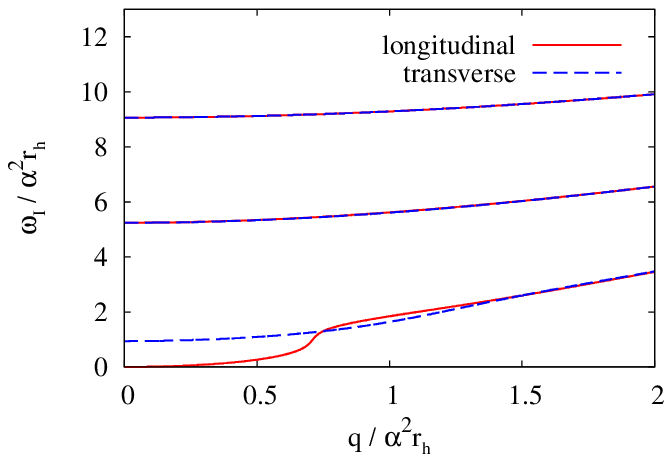, height=7.14cm, width=4.97cm,
angle=270}
\centering\epsfig{file=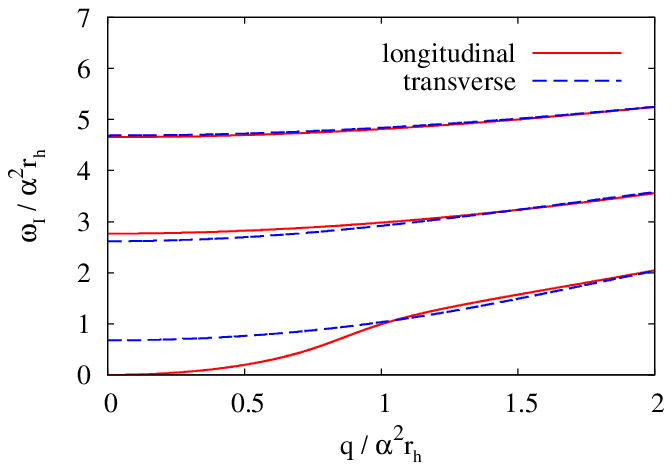, height=7.14cm, width=4.97cm,
angle=270} 
\caption{Dispersion relations of a special kind of purely damped
electromagnetic QNM of the rotating black string for two different values of
the rotation parameter:  $a\alpha=0.5$ (left) and $a\alpha=0.7$ (right). These are
special modes in the sense that they are not found for small rotation,
neither have similar counterparts in the static case. See also table
\ref{purelydamped}.}
\label{modospur2}
}
\TABLE{
\begin{tabular}{ccccc}
\hline 
\hline &\multicolumn{2}{c}{$a\alpha=0.5$} & 
\multicolumn{2}{c}{$a\alpha=0.7$}
\\ \hline
 $n_{\ss{S}}$&Longitudinal & Transverse & 
Longitudinal & Transverse\\
\hline 
0 & 0       & 0.940071 &   0     & 0.677287\\
1 & 5.24188 & 5.23989 & 2.76648 & 2.61773 \\
2 & 9.06564 & 9.06564 & 4.65369 & 4.68590 \\
3 & 12.8122 & 12.8122 & 6.60254 & 6.59832 \\
4 & 16.5355 & 16.5355 & 8.51922 & 8.51901 \\
 \hline\hline
\end{tabular}
\caption{Five different overtones of purely damped QNM for
longitudinal and transverse electromagnetic perturbation sectors with
$a\alpha=0.5$ and $a\alpha=0.7$, and with $\mathfrak{q}=\mathfrak{m}=0$. See
also figure~\ref{modospur2}. \label{purelydamped}} }

\subsection{Purely damped quasinormal modes - fast rotation}

It is worth saying that for small nonzero values of the rotation
parameter it was not found other purely damped modes in any region of
the spectrum (for finite values of frequencies and wavenumbers, with
$\mathfrak{m} =0$). Meanwhile, for higher values of the rotation parameter
we have also found a special series of purely damped modes, both for 
longitudinal and transverse perturbation sectors,  which do not
present saturation points. They are special modes in the sense
that they are not found for small values of the rotation parameter, neither
have similar counterparts in the static case. We show a few of such modes
for $a\alpha=0.5$ and $a\alpha=0.7$ in figure \ref{modospur2}.
We note that, for high values of the rotation parameter, the longitudinal
and transverse quasinormal frequencies have very close values. These
results indicate the isospectrality of the purely damped modes in high
rotation. The same effect seems to happen for the other class of purely
damped modes found in the slow rotating case, namely, the modes that exist at
high wavenumbers and high frequencies (cf. the upper lines in figure
\ref{modospur}). To confirm the isospectrality of the modes of figure
\ref{modospur2}, we list the
values of the quasinormal frequencies for the first five of these
modes calculated at zero wavenumbers ($\mathfrak{q}=\mathfrak{m}=0$) in
table \ref{purelydamped}.

\subsection{Ordinary quasinormal modes}

As in the case of static black holes, the electromagnetic perturbations of
rotating black strings also present a group of ordinary (or regular) QNM
characterized by having nonzero frequencies (both real and
imaginary parts) in the zero wavenumber limit, $\mathfrak{m}\rightarrow0$ and
$\mathfrak{q}\rightarrow0$. Here we investigate these modes for longitudinal
and transverse sectors considering a few different values of the rotation
parameter ($a\alpha=0.1,\;0.2,\;0.5$ and $0.7$). 

In table \ref{taba01} we present the quasinormal frequencies for the first
four transverse and longitudinal regular modes calculated with
$a\alpha=0.1$ and $\mathfrak{q}=\mathfrak{m}=0$.

\TABLE{
\begin{tabular}{ccccc}
\hline 
\hline &\multicolumn{2}{c}{Longitudinal} & 
\multicolumn{2}{c}{Transverse}
\\ \hline
 $n$&$\quad\;\;\mathfrak{w}_{\ss{R}}\quad\;\;$ &
$\quad\;\;\mathfrak{w}_{\ss{I}}\quad\;\;$ &
$\quad\;\;\mathfrak{w}_{\ss{R}}\quad\;\;$ &
$\quad\;\;\mathfrak{w}_{\ss{I}}\quad\;\;$\\
\hline 
1 & 1.49581 & 4.13330 & 2.25027 & 5.15701 \\
2 & 2.98987 & 6.21144 & 3.71623 & 7.28524 \\
3 & 4.43191 & 8.37247 & 5.13900 & 9.46958 \\
4 & 5.83910 & 10.5744 & 6.53342 & 11.6853 \\
 \hline\hline
\end{tabular}
\caption{Ordinary QNM for longitudinal and transverse
electromagnetic perturbations with $a\alpha=0.1$ and
$\mathfrak{q}=\mathfrak{m}=0$.}
\label{taba01}}

In figure \ref{regular1} we plot the dispersion relations of the first
($n=1$) longitudinal and transverse ordinary QNM for $a\alpha=0.1$. The
graphs on the left hand side of the figure show real parts of the
frequency and the graphs on the right hand side show the imaginary parts of
the frequency.
It can be seen that, for both perturbation sectors, the imaginary parts of
the frequency are not symmetric with respect to the wavenumber
$\mathfrak{m}$. It is noteworthy that the symmetry with respect to the
wavenumber $\mathfrak{q}$ is preserved, so we plot
only the positive values of wavenumber $\mathfrak{q}$.
\FIGURE{
\centering\epsfig{file=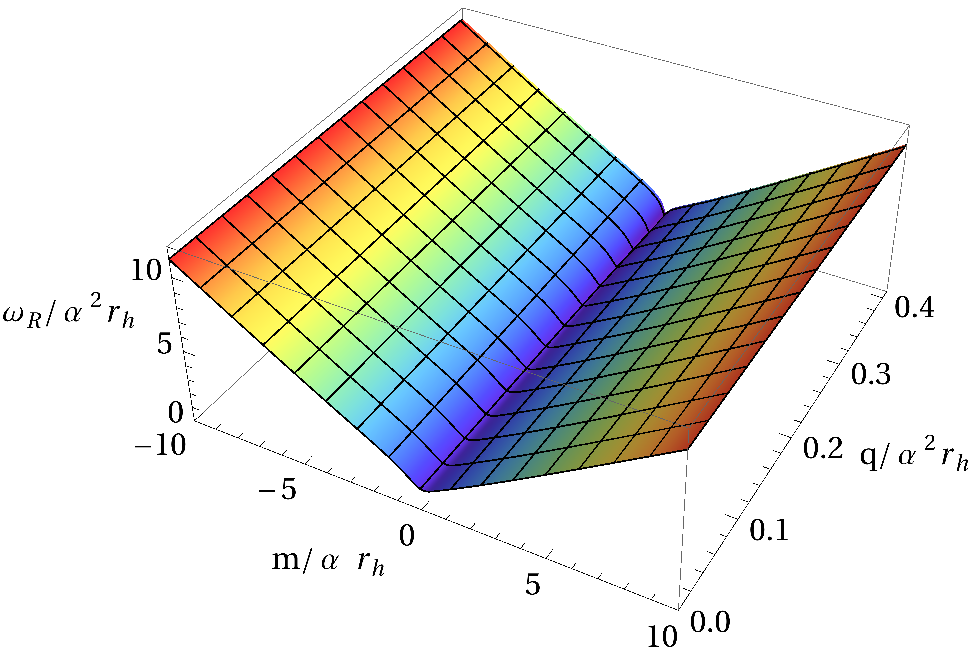,height=4.229cm,width=6.3625cm}
\centering\epsfig{file=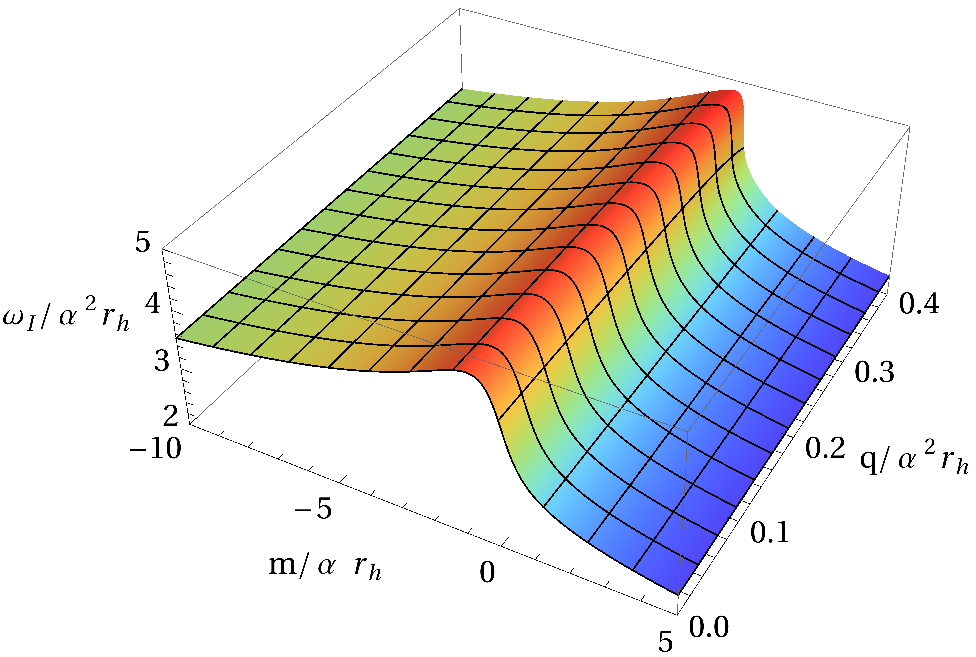,height=4.229cm, width=6.3625cm}
\centering\epsfig{file=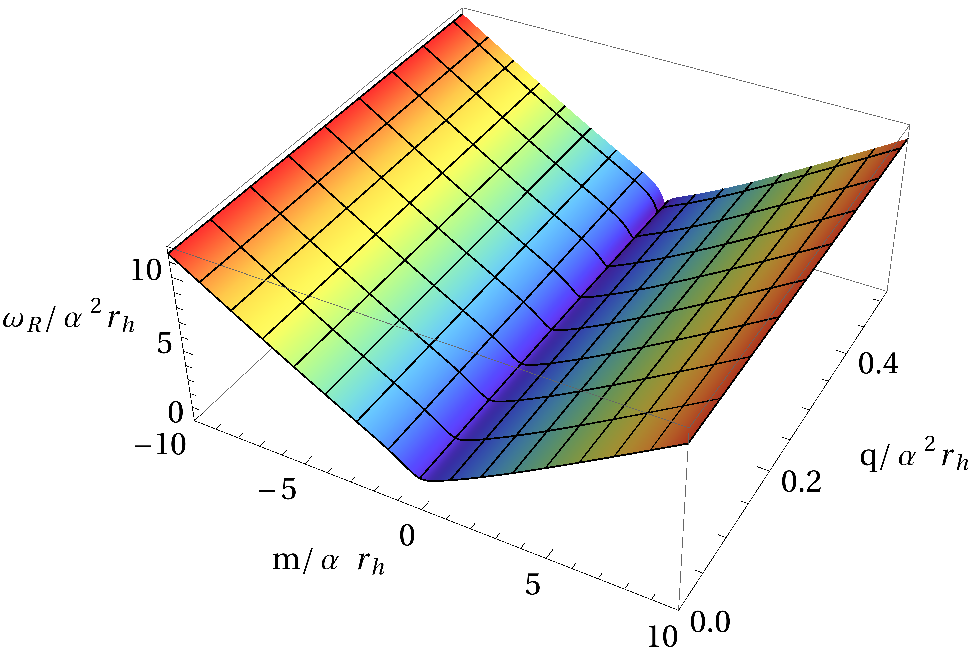,height=4.229cm,width=6.3625cm}
\centering\epsfig{file=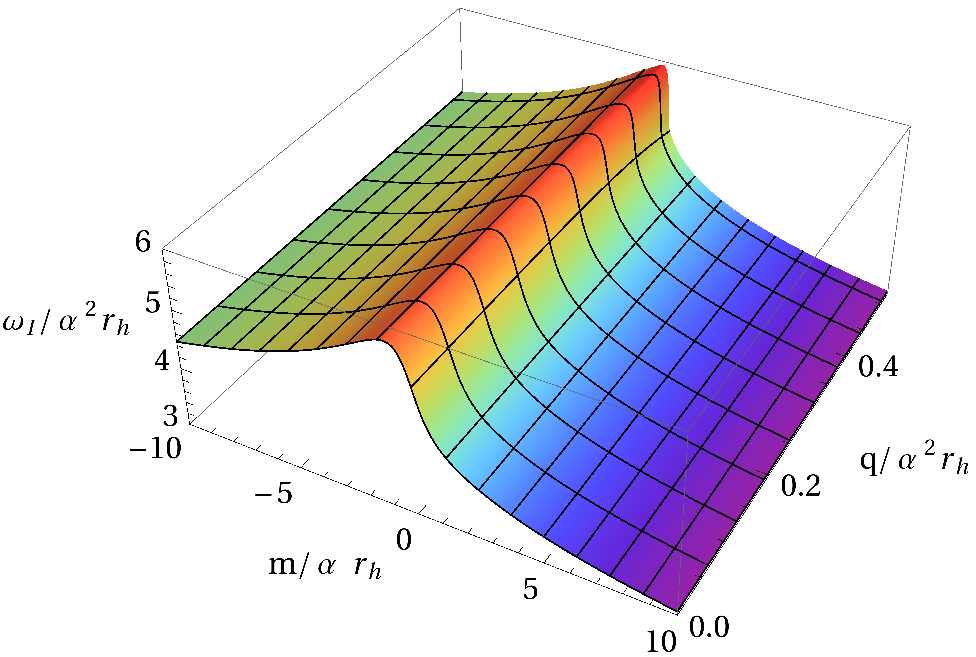,height=4.229cm, width=6.3625cm}
\caption{Dispersion relations of the first ($n=1$) longitudinal (top) and
transverse (bottom) ordinary QNM in the case $a\alpha=0.1$. The real parts
of the frequency are in the graphs on the left, and the imaginary parts of
the frequency are in the graphs on the right.}
\label{regular1}}
It is seen that the imaginary parts of the frequency decrease with the
wavenumber $\mathfrak{m}$, while the real parts increase. The imaginary
parts of the frequency decrease faster for positive 
values of the wavenumber $\mathfrak{m}$ than for negative values. Although
graphically the real parts of the frequency seem to be symmetric, the values
of the frequency for symmetric values of $\mathfrak{m}$ are different. For
example, the frequency of the longitudinal quasinormal
mode at $(\mathfrak{q},\,\alpha \mathfrak{m})=(0.2,\, +6.0)$  is
$\mathfrak{w}_{\ss{R}}=6.76969$, while at $(\mathfrak{q},\,
\alpha \mathfrak{m})=(0.2,\, -6.0)$  it is $\mathfrak{w}_{\ss{R}}=7.34660$.
The difference is easily understood as Doppler shift.

The quasinormal frequencies of a few ordinary modes found with 
$\mathfrak{q}=\mathfrak{m}=0$, for
$a\alpha=0.2,\;0.5$ and $0.7$, are shown in table \ref{tabqm0b}.
\TABLE{
\begin{tabular}{ccccccc}
\hline 
\hline\multicolumn{7}{c}{Longitudinal sector}
\\\hline & \multicolumn{2}{c}{$a\alpha=0.2$} & 
\multicolumn{2}{c}{$a\alpha=0.5$}& 
\multicolumn{2}{c}{$a\alpha=0.7$}
\\ \hline
$\;\;n\;\;$ &
$\quad\;\;\mathfrak{w}_{\ss{R}}\quad\;\;$ &
$\quad\;\;\mathfrak{w}_{\ss{I}}\quad\;\;$ & $\quad\;\;
\mathfrak{w}_{\ss{R}}\quad\;\;$ &
$\quad\;\;\mathfrak{w}_{\ss{I}}\quad\;\;$& $\quad\;\;
\mathfrak{w}_{\ss{R}}\quad\;\;$ &
$\quad\;\;\mathfrak{w}_{\ss{I}}\quad\;\;$ \\
\hline 
1 & $1.89070$ & 3.83011 & $0.698883$ & 1.41673 & 0.779558 &
1.28190 \\
2 & $3.37595$ & 6.00354 & $2.26543$ & 3.90299 & $2.70200$ &
4.68020\\
3 & $4.79244$ & 8.28380 & $3.95790$ & 6.85402 & $4.80012$ &
8.31405 \\
4 & $6.17560$ & 10.6313 & $5.67558$ & 9.83043 & $6.91203$ &
11.9720 \\
\hline\hline
\multicolumn{7}{c}{Transverse sector}
\\\hline
 & \multicolumn{2}{c}{$a\alpha=0.2$} & 
\multicolumn{2}{c}{$a\alpha=0.5$}& 
\multicolumn{2}{c}{$a\alpha=0.7$}
\\ \hline
$\;\;n\;\;$ &
$\quad\;\;\mathfrak{w}_{\ss{R}}\quad\;\;$ &
$\quad\;\;\mathfrak{w}_{\ss{I}}\quad\;\;$ &
$\quad\;\;\mathfrak{w}_{\ss{R}}\quad\;\;$ &
$\quad\;\;\mathfrak{w}_{\ss{I}}\quad\;\;$&
$\quad\;\;\mathfrak{w}_{\ss{R}}\quad\;\;$ &
$\quad\;\;\mathfrak{w}_{\ss{I}}\quad\;\;$ \\
\hline 
1 & $1.11157$ & 2.82061 & $1.49788$ & 2.52890 & $1.66774$ &
2.89436 \\
2 & $2.64477$ & 4.89856 & $3.10140$ & 5.37862 & $3.74734$ &
6.49058\\
3 & $4.09020$ & 7.13378 & $4.81471$ & 8.33942 & $5.85513$ &
10.1414 \\
4 & $5.48647$ & 9.45034 & $6.53797$ & 11.3241 & $7.96969$ &
13.8039 \\
\hline\hline
\end{tabular}
\caption{The longitudinal and the transverse frequencies of the first four
ordinary QNM for $a\alpha=0.2$, $a \alpha =0.5$, and $a\alpha =0.7$,
calculated with $\mathfrak{q}=\mathfrak{m}=0$.} \label{tabqm0b}}
The dispersion relations for these quasinormal modes present similar behavior
as those showed in figure \ref{regular1}, and therefore they will not be
presented here.

From the point of view of AdS/CFT correspondence, the rotating black string
studied here is holographically dual to a CFT plasma moving with respect to
the preferred rest frame introduced by the topology. Therefore, an observer
at AdS boundary (CFT side) can see a wave propagating along the direction of
the motion of the plasma (with wavenumber $+\mathfrak{m}$), and another wave
propagating along the opposite direction (with wavenumber $-\mathfrak{m}$).
This justifies the fact that the numerical results for the frequency show a
symmetry breaking with respect to the wavenumber $\mathfrak{m}$. In fact,
this break of symmetry happens in both the longitudinal and the transverse
sectors of the electromagnetic fluctuations, and it is due to kinematic
effects like the Doppler shift.

The overall effect of the rotation in the quasinormal frequencies of AdS
black strings can be analyzed through the graphs of figure
\ref{a_varia}. We have chosen to show here the
fundamental quasinormal modes for each perturbation sector because these
modes are important from the point of view of AdS/CFT correspondence.
The graphs of figure \ref{a_varia} show the evolution of the frequency as
a function of the rotation parameter for fixed wavenumber ($\mathfrak{m}=0$
and $\mathfrak{q}=0.1$).
\FIGURE{
\centering\epsfig{file=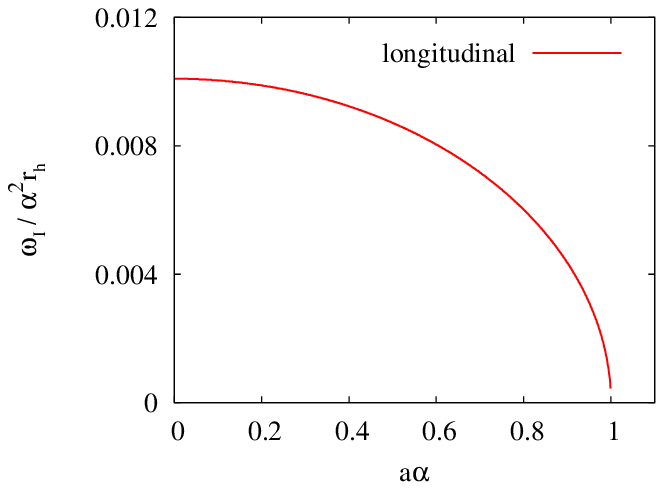, height=7.14cm,
width=4.97cm,angle=270}
\centering\epsfig{file=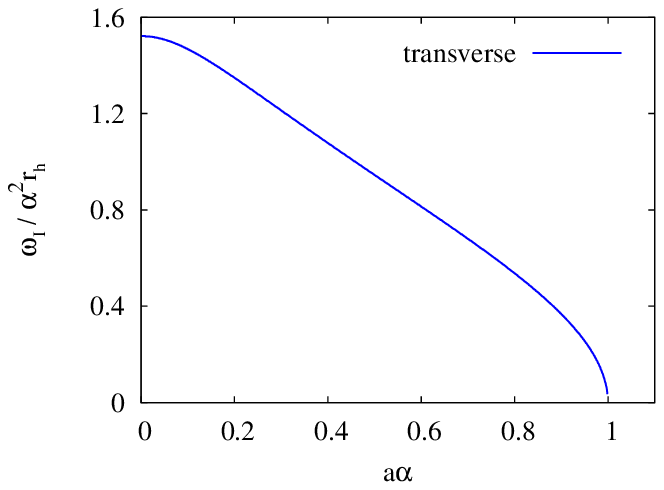, height=7.14cm, width=4.97cm,
angle=270}
\caption{Imaginary parts of the quasinormal frequencies as a function of
the rotation parameter with fixed wavenumber ($\mathfrak{m}=0$ and
$\mathfrak{q}=0.1$). On the left, the longitudinal (hydrodynamic) mode and
on the right the first transverse (purely damped) mode.}
\label{a_varia}}
It is seen that the rotation implies in a contraction of the  imaginary
parts of the fundamental quasinormal frequencies for both perturbation
sectors.
Based on the AdS/CFT correspondence, this behavior means that the rotation
implies in a dilation of the thermalization time of the dual CFT plasma, in
agreement with our analytical results (cf. eq. \eqref{tau}).

\section{Final comments and conclusion}\label{secfinal}

In this work the previous study of electromagnetic perturbations of static
AdS black holes was extended to AdS rotating black strings
\cite{Lemos:1994xp}. The
electromagnetic perturbation equations were obtained in a covariant
form and the resulting expressions imply that the QNM frequencies and
wavenumbers of the rotating black string are related to the QNM
frequencies and wavenumbers of the static black string through
simple expressions (cf. equations \eqref{boost2}).

The longitudinal hydrodynamic quasinormal mode of the electromagnetic
perturbations of the rotating black strings was found numerically, by
directly solving the perturbation equations, and analytically through
relations \eqref{boost2}. Differently from the static case, this diffusive
mode has quasinormal frequency with nonzero real part, which can be
interpreted as a convection term that emerges due to the motion of fluid in
relation to the preferred inertial frame introduced by the topology. The
hydrodynamic quasinormal mode presents a new diffusion coefficient for the
corresponding fluid, $D=\bar{D}/\gamma$, that depends on the rotation
parameter and implies a contraction of the imaginary parts of the
frequencies. 

The non-hydrodynamic QNM were also obtained and some of their dispersion
relations have been presented. In comparison to the static case, the main
new characteristic is a break of symmetry in the dispersion relations under
inversion of the wavenumber component in the direction of the rotation
($\mathfrak{m}\rightarrow -\mathfrak{m} $), while the symmetry with respect
to the wavenumber perpendicular to the rotation direction ($\mathfrak{q}$) is
preserved. This symmetry breaking in the rotation direction can be associated
to Doppler shifts of the frequencies and wavenumbers.

Both analytically and numerically we observed that rotation implies in a
decreasing of the values of the imaginary parts of the quasinormal
frequencies for both of the perturbation sectors, and hence, following the
AdS/CFT duality, the rotation implies in dilation of the thermalization times
of the CFT plasma.

We have also found a special class of electromagnetic purely damped modes of
the rotating black string which exists only for modes propagating along the
perpendicular direction to the rotation ($\mathfrak{m}=0$).
More study is necessary in order to find possible interpretations of this
special family of quasinormal modes.

\section{Acknowledgments}

This work is partly supported by Funda\c{c}\~ao de Amparo
\`a Pesquisa do Estado de S\~ao Paulo (FAPESP, Brazil).
VTZ thanks Conselho Nacional de
Desenvolvimento Cient\'\i fico e Tecnol\'ogico (CNPq, Brazil) for financial
support.

\end{document}